\documentclass[preprint,showpacs,preprintnumbers,amsmath,amssymb,showkeys]{revtex4}
\def\texpsfig#1#2#3{\vbox{\kern #3\hbox{\includegraphics{#1}\kern #2}}\typeout{(#1)}}

\input epsf   

\usepackage{graphicx}
\usepackage{dcolumn}
\usepackage{bm}

\newcommand{\snl}{\stackrel{(0)}{<}}
\newcommand{\snr}{\stackrel{(0)}{>}}
\newcommand{\snra}{\stackrel{(1)}{>}}
\newcommand{\snrb}{\stackrel{(2)}{>}}

\begin{document}

\draft

\title{A Variational Perturbation Approximation Method \\ in Tsallis Non-Extensive Statistical Physics}

\author{Wen-Fa Lu}
   \email{wenfalu@online.sh.cn}
\affiliation{ \ Institute for Theoretical Physics, Department of
Physics, Shanghai Jiao Tong University, Shanghai 200030, the
People's Republic of China }
\date{\today}

\begin{abstract}
For the generalized statistical mechanics based on the Tsallis
entropy, a variational perturbation approximation method with the
principle of minimal sensitivity is developed by calculating the
generalized free energy up to the third order in variational
perturbation expansion. The approximation up to the first order
amounts to a variational approach which covers the variational
method developed in Phys. Rev. Lett. 80, 218 (1998) by Lenzi $et\;
al$, and the approximations up to higher orders can systematically
improve variational result. As an illustrated example, the
generalized free energy for a classical harmonic oscillator
(considered in the Lenzi's joint work) are calculated up to the
third order, and the resultant approximations up to the first,
second, and third orders are numerically compared with the exact
result.
\end{abstract}

\pacs{ 05.70.Ce; 05.10.-a; 05.30.Ch}

\keywords{non-extensive effect, Tsallis entropy, variational
perturbation theory}

\maketitle

\section{Introduction}
\label{1}

Tsallis non-extensive statistical physics (TNESP) is one of
theories for generalizing the Boltzmann-Gibbs statistical
mechanics and thermodynamics \cite{1}. Its formalism is based on
the Tsallis entropy with a parameter $q$, the index of
non-extensiveness for a system. It takes the conventional
statistical mechanics as its special case of $q\to 1$. A distinct
feature of it is the power-law distributions ($q\not= 1$) instead
of the exponential law in the conventional statistical mechanics.
Since Tsallis' pioneered work in 1988 \cite{2}, the TNESP has
greatly developed, and successfully been used for investigating
many systems with a long-range interaction, long-time memory, or
fractal structured space-time \cite{1}, in which the non-extensive
effect exists \cite{1}.

For a new physical theory, to develop basic and analytic
approximation tools for the TNESP should be a fundamental task for
developing and employing it. To finish this task is particularly
necessary  for the TNESP because to perform calculation with the
Tsallis statistics is usually difficult due to the presence of the
power-law distributions. Since the TNESP with $q=1$ is reduced to
the conventional statistical mechanics, it will be natural to
generalize variational method and perturbational theory, the two
basic approximation methods in the conventional statistical
mechanics \cite{3}, to the TNESP. In fact, some progress has been
made in this direction. Early in 1993, employing concavity
properties of Tsallis entropy, Plastino and Tsallis established a
generalized Bogoliubov inequality and accordingly developed a
scheme of variational approximation for the free energy in the
generalized canonical ensemble \cite{4}. Later in 1998, Lenzi
$et\; al.$ generalized the perturbation method by deriving
approximate expressions of the generalized free energy (GFE) up to
the second order and obtained, through analyzing the approximate
expression up to the second order, another generalized Bogoliubov
inequality which affords a different variational method \cite{5}.
Recently, Mendes $et \; al.$ gave a comparative study on the
aforementioned variational methods \cite{6}, and illustrated that
the variational method in Ref.~\cite{5} gives a better
approximation for $q<1$ and is easier to perform than the
variational method in Ref.~\cite{4}, albeit the latter can lead to
a better approximation than the former for the case of $1<q<2$.

Nevertheless, the usefulness of both variational and perturbation
methods is limited. The variational approximate scheme in
Ref.~\cite{4} is valid only for the case of $q<2$, and the
variational approximate scheme in Ref.~\cite{4} works only for the
case of $q>0$. As for the perturbation method, the $n$th-order
perturbation approximation can be used only when $q>(1-{\frac
{1}{n}})$. Moreover, as is well known in the conventional
statistical mechanics, to control the approximate accuracy of
variational methods is not straightforward, and the perturbation
method is valid only for an exactly soluble system with a really
small perturbation. Hence, for systematically improving
variational method and extending the valid range of $q$, it is
worthwhile developing new approximate approaches.

In the conventional statistical mechanics and some other branches
of physics, a variational perturbation idea \cite{7} which
collects merits and overcomes drawbacks of both the perturbation
and the variational methods has developed and now becomes a
powerful tool called variational perturbation theory \cite{8}
(sometimes nominated other names.). The author or with his
collaborators also developed some schemes for the variational
perturbation theory \cite{9}(Refs.~\cite{8,9} have briefly
introduced references on the schemes for variational perturbation
theory). In the present paper, we intend to generalize the
variational perturbation theory to the TNESP. In the generalized
canonical ensemble, we will consider an expansion for the GFE
analogous to that in Ref.~\cite{5} but with artificially
introduced some auxiliary parameter $\mu$ and an expansion index
$\epsilon$, and derive the expressions of the first four terms in
the expansion series which will lead to approximations for the GFE
up to the third order. From the truncated expressions of the GFE
at the first, second and third orders (with taking $\epsilon=1$),
the auxiliary parameter $\mu$ will be determined according to a
principle of minimal sensitivity (PMS) \cite{10} at the first,
second and third orders, respectively. The value of $\mu$
determined at some order entering the truncated expression of the
GFE at the same order produces the approximation for the GFE up to
the same order. One will see that the adoption of the PMS is
crucial to the variational perturbation approximation (VPA)
method. It is the PMS that guarantees the meaningfulness,
non-perturbational nature and effectiveness of the VPA results. To
illustrate the method, we will consider a classical harmonic
oscillator, which was used in Ref.~\cite{5}, and calculate the GFE
up to the third order in the variational perturbation expansion.
Numerical comparisons between the first-, second- and third-order
and exact results are made and indicate that the VPA method
provides a better approximation than the variational method in
Ref.~\cite{5}.

Nowadays, owing to the so-called ``normalization" problem of
Tsallis non-extensive thermostatistics, there have existed four
versions for the TNESP \cite{2,11,12,13}. The first version with
the usual constraint on the internal energy in Ref.~\cite{2} was
used, for some special systems, only a couple of times in the
past, and the second version with the non-normalized constraint on
the internal energy in Ref.~\cite{11} has been intensively studied
and used, and furthermore the aforementioned variational and
perturbation methods were developed in this version. The third
version with the normalized constraint on the internal energy in
Ref.~\cite{12} is a satisfactory version, but is very complicated
for performing because there exist implied relations between
relevant quantities, and so the fourth version with the
non-normalized constraint on centered operators appears to
unentangle the problem on the implied relations in the third
version \cite{13}. Happily, the four versions can be easily
derived from just any one of them \cite{12,14}. Moreover, we
notice that the second version has been applied to many systems
with providing satisfactory theoretical and/or experimental
results \cite{1,12}. Hence, here we will use the second version
worked out in Ref.~\cite{11} to perform the VPA method.

Next, the Tsallis statistical mechanics with the canonical
ensemble and the perturbation and variational methods in
Ref.~\cite{5} will briefly be introduced for the convenience of
our later investigations. In Sect. III, the VPA method will be
stated and truncated expressions of the expansion series of the
GFE in the second version in Ref.~\cite{11} for a system will be
derived up to the third order. Sect. IV will contribute to
investigating the classical harmonic oscillator, and conclusions
will be made in Sect. V.

\section{Tsallis Statistical Mechanics}
\label{2}

For a system with a non-extensive index $q$, the Tsallis entropy
is defined as \cite{2}
\begin{equation}
S_q=k {\frac {1-\sum^W_{i=1} p_i^q}{q-1}} ,
\end{equation}
where $k$ is the Boltzmann constant and $W$ is the total number of
microscopic possibilities $p_i$ of the system. Note that $q$ is a
real number and characterizes the degree of non-extensiveness.
When $q=1$, Eq.(1) leads to the usual entropy $S_1 =-k\sum^W_{i=1}
p_i\ln p_i$. The three cases of $q<1$, $q=1$ and $q>1$
characterize superextensivness, extensiveness and subextensiveness
of systems, respectively. Based on the Tsallis entropy, a
generalized equilibrium thermostatistics can be established by
making $S_q$ extremal with appropriate constraints present. For
the canonical ensemble, four possible candidates exist for the
constraints and hence lead to the aforementioned four versions for
Tsallis statistical mechanics. This section will use the following
non-normalized constraint on the internal energy $U_q$ \cite{11}:
\begin{equation}
\sum^W_{i=1} p_i^q E_i =U_q,
\end{equation}
where $E_i$ is the $i$th eigenvalue of the Hamiltonian of the
system. (The corresponding constraint used in the first version of
Tsallis statistical mechanics is $\sum^W_{i=1} p_i E_i =U_q$, and
the one in the third version is the normalized one $\sum^W_{i=1}
p_i^q E_i/\sum^W_{i=1} p_i^q  =U_q$.) Thus, making $S_q$ extremal
with present the constraint Eq.(2) and the normalization property
of the probability yields the following power-law distribution
\cite{11}
\begin{equation}
p_i=p_i(E_i)=p_i(\beta^*)={\frac {[1-(1-q)\beta^*
E_i]^{1/(1-q)}}{Z_q(\beta^*)}}
\end{equation}
with the generalized partition function
\begin{equation}
Z_q(\beta^*)=\sum^W_{i=1}[1-(1-q)\beta^* E_i]^{1/(1-q)},
\end{equation}
where $\beta^*$ is the Lagrange multiplier associated with the
internal-energy constraint, Eq.(2). In the case of $q<1$, the
summation in the distribution Eq.(2) will be cut off for those
energy eigenvalues higher enough to give negative probabilities.
It is easy to verify that when $q\to 1$, the power-law
distribution Eq.(2) tends to the conventional exponential-law
distribution.

From Eqs.(2), (3) and (4), one can easily read
\begin{equation}
U_q=-{\frac {\partial}{\partial \beta^*}}{\frac
{Z_q^{1-q}-1}{1-q}}=-{\frac {\partial ln_q (Z_q)}{\partial
\beta^*}},
\end{equation}
where the $q$-logarithm function $\ln_q (x)\equiv
(x^{1-q}-1)/(1-q)$ is a generalization of the usual logarithm
function. Through Legendre transform on $\ln_q (Z_q)$, which
depends on $\beta^*$, one can find the relation $S_q=k(ln_q
(Z_q)+\beta^* U_q)$. Note that the relation between the
generalized internal energy and Tsallis entropy takes the same
form as in the conventional statistical mechanics. In fact,
introducing $t\equiv 1/(k \beta^*)$, and defining generalized
thermodynamic function in the same way as that in the conventional
thermodynamics, one can find the same Legendre structure as in the
conventional thermodynamics. For example, one can have the GFE
\begin{equation}
F_q\equiv U_q-t S_q=F_q(\beta^*)= -{\frac {1}{\beta^*}}{\frac
{Z_q^{1-q}-1}{1-q}}=-{\frac {1}{\beta^*}}\ln_q (Z_q) \;.
\end{equation}
Using the basic limit formula $\lim_{x\to 0}(1+ax)^{{\frac
{1}{x}}}=e^a$, one can verify that the above generalized
thermodynamic function and partition function are reduced to the
usual ones in the limit of $q\to 1$.

From the above, one can see that although the usual
thermodynamical Legendre structure remains valid in Tsallis
statistical mechanics, it is evident that the original calculation
techniques can not be directly borrowed into the generalized
theory. One has to design calculation techniques for the
generalized theory. As was stated in the introduction, a united
presentation of the perturbation and variational methods for the
TNESP has been given in Ref.~\cite{5}. Next, for convenience of
later contrast, we give a brief introduction on them in the
notations here.

Assume that the Hamiltonian can be written as
\begin{equation}
H=H_0+\lambda H_I ,
\end{equation}
where, $H_0$ is the Hamiltonian of a soluble model, $\lambda H_I $
is small enough so that it can be considered as a perturbation on
$H_0$, and $\lambda$ is the perturbation parameter. The GFE of the
system with $H$, $F_q(\lambda)$, is a function of $\lambda$ and
can be expanded as
\begin{equation}
F_q(\lambda)=F_q^{(0)}+\lambda F_q^{(1)} +{\frac
{\lambda^2}{2!}}F_q^{(2)} + \cdots
\end{equation}
with $F_q^{(1)}={\frac {\partial F_q(\lambda)}{\partial
\lambda}}|_{\lambda=0}$, $F_q^{(2)}={\frac {\partial^2
F_q(\lambda)}{\partial \lambda^2}}|_{\lambda=0}$, $\cdots$. Eq.(8)
is a perturbation expansion series of $F_q(\lambda)$ in powers of
$\lambda$, and to truncate it at some order, for example $n$th
order, can provide an approximation to $F_q(\lambda)$ up to the
$n$th order. In Ref.~\cite{5}, approximate expressions for
$F_q(\lambda)$ was derived up to the second order. Through
analyzing the approximation expression of $F_q(\lambda)$ up to the
second order, Ref.~\cite{5} obtained the generalized Bogoliubov
inequality
\begin{equation}
F_q\le F_q(0)+\sum^W_{i=1}p_i^q (E_i^{(0)})\snl i|\hat{H}_I|i\snr,
\end{equation}
where the symbol $\hat{O}$ is the operator corresponding to the
physical quantity $O$ and $|i\snr$ (and $\snl i|$) is the $i$th
eigenstate for $\hat{H}_0$. This inequality stems from the
approximation expression of $F_q(\lambda)$ up to the first order
in the perturbation expansion, and affords a variational method.

According to Ref.~\cite{5}, the approximate expression of
$F_q(\lambda)$ up to the $n$th order is valid only for the case of
$q>1-{\frac {1}{n}}$ because the derivation of the approximate
expressions requires the interchangeability between the sum over
the quantum numbers and the derivative with respect to $\lambda$.
As was stated below Eq.(4), when $q<1$, the sum in Eq.(4) and,
accordingly, Eq.(6) has a cutoff and so its upper limit depends
upon $\lambda$, leading to the above limitation on the resultant
expressions in Ref.~\cite{5}. Therefore, the variational method in
Ref.~\cite{5} can be used only for $q>0$.

We stop here for introducing the perturbation and variational
methods in Ref.~\cite{5}. Next, we turn to our investigation in
the present paper.

\section{Variational Perturbation Approximation Scheme}
\label{3}

For simplicity of the notation, $k=1$ will be taken from now on.

To develop the VPA method, we consider calculating the GFE $F_q$
(Of course, the method can be used to calculate other generalized
thermodynamic potentials). We begin with modifying the Hamiltonian
$H$. Firstly, differently from Eq.(7) in the perturbation method,
we add the zero term $H_0(\mu)-H_0(\mu)$ to $H$ and write it as
\begin{equation}
H=H_0(\mu)+H_I(\mu)
\end{equation}
with $H_I=H-H_0(\mu)$ and $\mu$ an auxiliary parameter. In
Eq.(10), $H_0(\mu)$ is the Hamiltonian of some exactly soluble
model which is originally not included in $H$, and the exact
solubility of the model is not affected by values of the auxiliary
parameter $\mu$. That is to say, for a value of $\mu$, the energy
eigenvalues $E_n^{(0)}$ (non-degenerate, for simplicity),
eigenstates $|n\snr$, the probability $p_q(E_n^{(0)})$, the
generalized partition function $Z_{q,\; 0}$ and the GFE $F_q^0$
for the model with $H_0(\mu)$ are exactly known. Secondly, an
artificial parameter $\epsilon$ is inserted as a factor before
$H_I (\mu)$ in Eq.(10), that is, $H$ in Eq.(10) is now modified as
\begin{equation}
H_\epsilon=H_0(\mu)+\epsilon H_I(\mu) .
\end{equation}
Thus, we have a new system with the Hamiltonian $H_\epsilon$ which
becomes $H_0(\mu)$ and $H$ in the cases of $\epsilon=0$ and
$\epsilon=1$, respectively. Note that different from $H$ in
Eq.(10), $H_\epsilon$ depends on $\mu$ for $\epsilon\not=1$. The
above modification provides a possibility that we calculate $F_q$
for the original system through considering the new system.

For the system with $H_\epsilon$, suppose that the set of energy
eigenvalues are the spectra ${E_{n,\; \epsilon}}$, and the
corresponding eigenfunctions is $|n,\epsilon>$.  When the system
is in thermal contact with a reservoir, $i.e.,$ when we adopt a
generalized canonical ensemble, we have the probability
distributions ${p_n(E_{n,\; \epsilon})}$, and can calculate the
GFE $F_q (\epsilon)$ in the presence of the constraint Eq.(2)
according to the definition and expression Eq.(6). In analogy to
Eq.(8), one can expand $F_q (\epsilon)$ as the following series
\begin{equation}
F_q(\epsilon)=F_q^{(0)}+\epsilon F_q^{(1)} +{\frac
{\epsilon^2}{2!}}F_q^{(2)} +
              {\frac {\epsilon^3}{3!}}F_q^{(3)} + \cdots
\end{equation}
with $F_q^{(k)}={\frac {\partial^k F_q(\epsilon)}{\partial
\epsilon^k}}\bigl |_{\epsilon=0}$ and $F_q^{(0)}=F_q(0)$. This is
only a formal expansion, and the parameter $\epsilon$ acts as an
expansion order index and needn't be small. Taking $\epsilon=1$ in
the last equation, we obtain an expansion series of $F_q$ for the
system with $H$ which is independent of $\mu$ if the resultant
series is not truncated.

Now we have a simple discussion on the expansion Eq.(12). Of
course, if $H_0(\mu)$ is naively one part of the original $H$ and
$H_I$ can be regarded as a perturbation on $H_0(\mu)$ (in this
case, $\mu$ is one of system parameters in $H$), then Eq.(12) with
$\epsilon=1$ is just the perturbation expansion series of $F_q$,
Eq.(8). If $H_0(\mu)$ is not a naive part of $H$, as we assumed in
Eq.(10), the truncated expression at the first order in $\epsilon$
can lead to a variational method, and substituting the value $\mu$
from the variational method into Eq.(12) with $\epsilon=1$ will
produce an expansion series of $F_q$ around the variational
result. In this case, if the resultant variational method can
provide a good approximation result for $F_q$, then it can be
presumed that the expansion of $F_q$ around the variational result
maybe afford a good non-perturbational approximation which will
improve variational results. This is the original version of the
variational perturbation theory, but the expansion series is
possibly divergent \cite{8}. If the variational method cannot
provide a good approximation result for $F_q$, the expansion of
$F_q$ around the variational result will make no senses. That is
to say, in general, the above expansion series Eq.(12) seems to be
useless.

However, as was mentioned in Sect.I, once the PMS is employed to
single out an appropriate value of $\mu $, Eq.(12) can afford a
presumably convergent, systematic non-perturbation approximation
method, a modern version of the variational perturbation theory
\cite{8}. In Eq.(12), truncating the expansion series at some
order in $\epsilon$ and then letting $\epsilon=1$, the resultant
truncated expression is a function of the parameter $\mu$, albeit
Eq.(12) with $\epsilon=1$ is independent of $\mu$. In principle,
the value of the truncated expression can possibly take any value
with variance of $\mu$, including the exact value of $F_q$. Since
$F_q$ is a constant in the space of $\mu$ (the exact $F_q$ is
independent of $\mu$), it should be believable that a requirement
of the truncated result varying most slowly with the parameter
$\mu$ can likely make the truncated result provide a most reliable
approximate result for the exact $F_q$. Employing this requirement
to determine the parameter $\mu$ is the main spirit of the so
called PMS \cite{10}. Generally, the curve of a function varies
more slowly near its extremum than in its slope part, and so a
simple realization of the PMS is to require the first derivative
of the truncated expression with respect to $\mu$ to be zero.
Thus, substituting the appropriate value of $\mu$ determined at
the truncated order into the corresponding truncated expression of
$F_q$ will provide a reasonable and reliable approximation for
$F_q$ up to the same order. Such a truncated result at some order
is the approximate result of $F_q$ up to the same order in the VPA
method.

In brief, the VPA method consists in only two crucial steps: one
is to formally expand the quantity $F_q$ in consideration with an
adjustable parameter $\mu$ entered, and the other is to determine
the value of $\mu$ from the truncated expression of $F_q$
according to the PMS. Now, we continue to perform the first step,
derive the first four terms in Eq.(12) (the derivations here are
similar to those in Ref.~\cite{5}), and simultaneously indicate
the VPA procedures up to the first, second and third orders,
respectively.

The first term in Eq.(12) is straightforwardly
\begin{equation}
F_q^{(0)}\equiv F_q^{(0)}(\mu)= -{\frac {1}{\beta^*}}{\frac
{{(Z_{q,\; 0}(\beta^*))}^{1-q}-1}{1-q}}
\end{equation}
with
\begin{equation}
Z_{q, \; 0}(\beta^*)=\sum^W_{n=1}[1-(1-q)\beta^*
E_n^{(0)}]^{1/(1-q)}.
\end{equation}

For deriving the other three terms, eigenstates of
$\hat{H}_\epsilon$ are needed, and so we first consider the
eigenequation, $\hat{H}_\epsilon|n,\epsilon>=E_{n,\;
\epsilon}|n,\epsilon>$. By mimicing Rayleigh-Schr\"odinger
perturbation theory, after substituting the expansions
\begin{equation}
|n,\epsilon>=|n\snr+\epsilon|n\snra+\epsilon^2|n\snrb+ \cdots
\end{equation}
and
\begin{equation}
E_{n,\; \epsilon}=E_n^{(0)}+\epsilon E_n^{(1)}+\epsilon^2
E_n^{(2)}+\cdots
\end{equation}
into the eigenequation, one can formally have
\begin{equation}
|n\snra=\sum_{m\not=n} {\frac {H_{I,\;
mn}}{E_n^{(0)}-E_m^{(0)}}}|m\snr \;,
\end{equation}
\begin{eqnarray}
|n\snrb&=&\sum_{m\not=n} \biggl [\sum_{l\not=n}{\frac
{H_{I,ml}H_{I,ln}}{
          (E_n^{(0)}-E_m^{(0)})(E_n^{(0)}-E_l^{(0)})}}-
        {\frac {H_{I,\; mn}H_{I,\; nn}}{(E_n^{(0)}-E_m^{(0)})^2}}\biggr ]|m\snr
    \nonumber \\       &&\hspace*{0.2cm}
           -{\frac {1}{2}}\sum_{m\not=n} {\frac {(H_{I,\; mn})^2}{(E_n^{(0)}
           -E_m^{(0)})^2}}|n\snr
\end{eqnarray}
and so on. Here, $H_{I,\; mn}= \snl m|\hat{H}_I(\mu)|n\snr$. By
the way, the expansion expressions, Eqs.(15) and (16), together
with the PMS can lead to a VPA method to solve Schr\"odinger
equation \cite{9}(PRD)\cite{15}.

For the second term of Eq.(12), $F_q^{(1)}={\frac {\partial
F_q(\epsilon)} {\partial \epsilon}}\bigl |_{\epsilon=0}$ can be
calculated as
\begin{equation}
F_q^{(1)}=\biggl [{\frac {\partial}{\partial \epsilon}}
                  {\frac {1-Z_{q,\; \epsilon}^{1-q}}{\beta^*(1-q)}}\biggr]\biggl
     |_{\epsilon=0}=Z_{q,\; \epsilon}^{-q} \sum_n [1-(1-q)\beta^* E_n]^{\frac {q}
     {1-q}}
     {\frac {\partial E_{n,\; \epsilon}}{\partial \epsilon}}\biggl|_{\epsilon=0}
\end{equation}
with $Z_{q,\; \epsilon}$ the generalized partition function for
the system with $H_\epsilon$. From Hellmann-Feynman theorem
${\frac {\partial E_{n,\; \epsilon}}{\partial
\epsilon}}=<\epsilon,n|{\frac {\partial \hat{H}_\epsilon}{\partial
\epsilon}} |n,\epsilon>$ \cite{16}, we have
\begin{eqnarray}
F_q^{(1)}&=&\biggl [Z_{q,\; \epsilon}^{-q} \sum_n [1-(1-q)\beta^*
E_{n,\; \epsilon}]
           ^{\frac {q} {1-q}}
     <\epsilon,n|\hat{H}_I|n,\epsilon>\biggr ]\biggl|_{\epsilon=0}
     \nonumber \\ &=&
     \sum_n p_q^q(E_n^0)H_{I,\; nn} = <H_I>_q^0 ,
\end{eqnarray}
which takes the same form as Eq.(5) in Ref.~\cite{5}. Thus, the
truncated expression for $F_q$ at the first order, $F_q^I(\mu)$,
is
\begin{equation}
F_q^I(\mu)=F_q^0+F_q^{(1)} ={\frac {1-Z_{q,\;
0}^{1-q}}{\beta^*(1-q)}}+<H_I>_q^0 \;.
\end{equation}
The right hand side of last equation has the same form as that in
the right hand side of the inequality, Eq.(9). $F_q^I(\mu)$ is a
function of $\mu$ and a different value of $\mu$ produces a
different approximation for $F_q$ which maybe have nothing to do
with the exact value of $F_q$. However, if the value of $\mu$ in
Eq.(21) is chosen from roots of the following condition
\begin{equation}
{\frac {\partial F_q^I(\mu)}{\partial \mu}}=0
\end{equation}
according to the PMS, then $F_q^I(\mu)$ in Eq.(21) would produce a
most reliable approximation for $F_q$ up to the first order. Since
$F_q$ has the same form as the right hand side of the inequality,
Eq.(9), the VPA procedure up to the first order for the case of
$q> 0$ is consistent with the variational method based on the
Bogoliubov inequality in Ref.~\cite{5}, and so the appropriate
root of Eq.(22) for $\mu$ makes $F_q^I(\mu)$ minimal. For the case
of $q\le 0$, an example in next section shows that the appropriate
root of Eq.(22) for $\mu$ also makes $F_q^I(\mu)$ minimal. This
suggests that the VPA procedure up to the first order is generally
a variational method which provides a reliable upper limit for the
exact $F_q$.

In order to give the truncated expression for the GFE at the
second order $F_q^{II}(\mu)=F_q^I(\mu) + F_q^{(2)}/2!$, we
calculate $F_q^{(2)}$ as follows. A straightforward
differentiation gives
\begin{eqnarray}
F_q^{(2)}&=&{\frac {\partial^2 F_q(\epsilon)}{\partial
\epsilon^2}}
          \biggl |_{\epsilon=0}
         =\beta^* q Z_{q,\; 0}^{q-1}\biggl [(<H_I>_q^0)^2  \nonumber   \\
         &&-\sum_n (p_q(E_n^{(0)}))^{2q-1}
           (H_{I,\; nn})^2\biggr ]+\sum_n (p_q(E_n^{(0)}))^q
           {\frac {\partial<\epsilon,n|H_I|n,\epsilon>}{\partial \epsilon}}
           \biggl|_{\epsilon=0} .
\end{eqnarray}
From Eq.(15), we have ${\frac {\partial|n,\epsilon>}{\partial
\epsilon}}\bigl|_{\epsilon=0}=|n\snra$. Consequently, Eq.(23)
becomes
\begin{eqnarray}
F_q^{(2)} &=&\beta^* q Z_{q,\; 0}^{q-1}\biggl
[(<H_I>_q^0)^2-\sum_n
            (p_q(E_n^{(0)}))^{2q-1} (H_{I,\; nn})^2\biggr ]
            \nonumber \\ &&+2\sum_n
            (p_q(E_n^{(0)}))^q \sum_{m\not=n}
           {\frac {|H_{I,\; nm}|^2}{E_n^{(0)}-E_m^{(0)}}} .
\end{eqnarray}
Formally, the right hand side of Eq.(24) can be rewritten as
Eq.(6) in Ref.~\cite{5}. According to the PMS, using the
reasonable root of the condition
\begin{equation}
{\frac {\partial F_q^{II}(\mu)}{\partial \mu}}=0
\end{equation}
as the value of $\mu$, one can give the approximation result for
$F_q$ up to the second order  from $F_q^{II}(\mu)$. The
approximation result up to the second order would improve the
approximation result up to the first order.

Similarly, one can calculate $F_q^{(3)}$. In the calculation, we
need the additional relation ${\frac
{\partial^2|n,\epsilon>}{\partial
\epsilon^2}}\bigl|_{\epsilon=0}=2|n>^{(2)}$, which can be easily
obtained from Eqs.(15), (17) and (18). Thus, $F_q^{(3)}$ can be
written as
\begin{eqnarray}
F_q^{(3)} &=& {\beta^*}^2 q(q+1) Z_{q,\;
0}^{2q-2}(<H_I>_q^0)^3-3{\beta^*}^2q^2
           Z_{q,\; 0}^{2q-2}\sum_n (p_q(E_n^{(0)}))^{2q-1}(H_{I,\; nn})^2<H_I>_q^0
           \nonumber \\ &&
           +{\beta^*}^2 q(2q-1)Z_{q,\; 0}^{2q-2}\sum_n (p_q(E_n^{(0)}))^{3q-2}
           (H_{I,\; nn})^3  \nonumber \\ &&
           +6\beta^* q Z_{q,\; 0}^{q-1}\sum_n
           \biggl [(p_q(E_n^{(0)}))^{q}<H_I>_q^0-(p_q(E_n^{(0)}))^{2q-1}
           H_{I,\; nn}\biggr ]
           \sum_{m\not=n}{\frac {|H_{I,nm}|^2}{E_n^{(0)}-E_m^{(0)}}}
               \nonumber  \\ &&
           +6\sum_n (p_q(E_n^{(0)}))^{q}  \biggl
            [\sum_{l\not=n}\sum_{m\not=n}{\frac {H_{I,nl}H_{I,lm}H_{I,nm}}
            {(E_n^{(0)}-E_m^{(0)})(E_n^{(0)}-E_l^{(0)})}}-
             \sum_{m\not=n} {\frac {|H_{I,nm}|^2 H_{I,\; nn}}
             {(E_n^{(0)}-E_m^{(0)})^2}} \biggr ].
\end{eqnarray}
Then one can have the truncated expression for $F_q$ at the third
order , $F_q^{III}(\mu)=F_q^{II}(\mu)+{\frac {1}{3!}}F_q^{(3)}$.
Furthermore, using the reasonable root of the condition
\begin{equation}
{\frac {\partial F_q^{III}(\mu)}{\partial \mu}}=0
\end{equation}
as the value of $\mu$, one can give the approximation result for
$F_q$ up to the third order  from $F_q^{III}(\mu)$.

In the same way, one can consider VPA results for $F_q$ up to
higher orders. Obviously, the VPA results up to various orders
constitute a sequence of the approximation results of $F_q$. Owing
to the employment of the PMS, the values of $\mu$ determined
depend upon the truncated orders, and the value of $\mu$ at a
given order is different from values of $\mu$ at other orders.
Presumably, it is the dependence of $\mu$ upon the truncated order
that makes the sequence of the VPA results of $F_q$ up to various
orders converge to the exact $F_q$. Although we cannot verify this
convergent property of the variational perturbation theory,
investigations in other fields have provided such a few examples
\cite{8,17}. Additionally, in the example of next section,
generally, the VPA result up to the third order approaches more
closely the exact $F_q$ than the approximations up to the lower
orders.

The second crucial step plays a vital role in the VPA method, and
one have to perform it very carefully. Sometimes, Eqs.(22), (25)
and (27) can not provide appropriate roots for $\mu$, and in those
cases one should render the second derivative of the truncated
expressions for $F_q$ with respect to $\mu$ zero to determine
$\mu$. This can be understandable because the curve of a function
varies more slowly near its knee than in its slope part (As a
matter of fact, when there exist both an extremum and a knee, one
should analyze behaviors of the truncated expressions as a
function of $\mu$ near the extremum and knee to determine $\mu$
according to the PMS). In case the second derivative condition can
not provide an appropriate root for $\mu$, too, then the VPA
method cannot provide an approximation result for $F_q$ up to the
truncated order in consideration, and one should further consider
VPA result up to the next order. Furthermore, when there exist
multi-roots of the vanishing derivatives for $\mu$, one should
choose to adopt the root near which the truncated result varies
most slowly with $\mu$. Next section, we will show how to
determine $\mu$ according to the PMS in a concrete example.

In concluding this section, we intend to emphasize one point. In
the above calculations, the interchange between the
differentiation with respect to $\epsilon$ and the sum over
eigenstates was involved. As was mentioned in last section, the
interchangeability between the differentiation and the sum leads
to a limitation for the variational and perturbation methods in
Ref.~\cite{5}. Nevertheless, for the VPA method developed here,
the relevant calculations are only formal calculations and the
resultant expressions are independent of $\epsilon$, and
presumably, we suggest that the VPA method needn't suffer those
limitations on the methods in Ref.~\cite{5}. At least, this point
yields no problem in the example of next section.

\section{A Classical Harmonic Oscillator}
\label{4}

For a one-dimensional classical harmonic oscillator with mass $M$
and an angular frequency $\omega$, the Hamiltonian is $H=p^2/(2
M)+M\omega^2 x^2/2$ with $x$ the coordinate and $p$ the momentum.
In a manner analogous to the one used for the quantum oscillator,
we can associate with the oscillator a continuous energy spectra
${E_n=\delta_o n}$, where $\delta_o$ is an arbitrary positive
constant with the dimension of energy and $n$ any positive real
number including $0$ \cite{12}. Then, from Eq.(4), the generalized
partition function can be written and calculated as
\begin{equation}
Z_q=\int_0^N[1-(1-q)\beta^* n\delta_o]^{1/(1-q)} dn={\frac
{1}{(2-q)\beta^*\delta_o}}
\end{equation}
with $N\to \infty$ for the case of $q>1$ and
$N=1/[(1-q)\beta^*\delta_o]$ for the case of $q<1$ (In this
section, we suppose that $\beta^*>0$.). The right hand side of
Eq.(28) is valid only for $q<2$ and $Z_q$ is divergent for $q>2$.
So, Eq.(6) leads to the GFE for the classical harmonic oscillator
\begin{equation}
F_q= -{\frac {1}{(1-q)\beta^*}}[({\frac
{1}{(2-q)\beta^*\delta_o}})^{1-q}-1].
\end{equation}
Last equation with $\delta_o=\hbar \omega$ is just Eq.(14) in
Ref.~\cite{5}.

On the other hand, for a particle with mass $M$ which moves in a
one-dimensional box, the Hamiltonian $H_0(L)=p^2/(2 M)+V_0$ with
$V_0=0$ for $|x|<{\frac {L}{2}}$ and $V_0\to \infty$ for $|x|\ge
{\frac {L}{2}}$. In analogy to what we did for the oscillator, we
can associate with the particle a continuous energy spectra
${E_n^{(0)}=\delta_b^2 n^2/(2 M L^2)}$, where $\delta_b$ is an
arbitrary positive constant with the dimension of action and $n$
any positive real number including $0$. So the generalized
partition function for the particle in the box can be written and
calculated as
\begin{eqnarray}
Z_{q,\; 0}=\int_0^N[1-(1-q)\beta^* {\frac {n^2\delta_b^2}{2 M
L^2}}]^{1/(1-q)} dn=\biggl \{
 \begin{array}{ll}
 {\frac {L}{2\delta_b}}\sqrt{{\frac {2 M}{(1-q)\beta^*}}}
                B({\frac {1}{2}},{\frac {2-q}{1-q}}) , & q<1
  \\ {\frac {L}{2\delta_b}}\sqrt{{\frac {2 M}{(q-1)\beta^*}}}
                B({\frac {1}{2}},{\frac {3-q}{2(q-1)}}) , & q>1
 \end{array}
\end{eqnarray}
with $N\to \infty$ for the case of $q>1$ and $N=L\sqrt{2
M/((1-q)\beta^*)}/\delta_b$ for the case of $q<1$. In Eq.(30),
$B(x,y)$ is the Beta function. Note that $Z_{q,\; 0}$ is
convergent only for the case of $q<3$. In the calculation of the
above equation, the formulae 8.380(1) and 8.380(3) in
Ref.~\cite{18} were employed. Owing to
$B(x,y)=\Gamma(x)\Gamma(y)/\Gamma(x+y)$ with $\Gamma(x)$ the gamma
function (8.384(1) in Ref.~\cite{18}), Eq.(30) with
$\delta_b=\hbar \pi$ is nothing but Eq.(11) in Ref.~\cite{5}. From
Eqs.(6) and (30), the GFE for the classical particle in the box is
easily calculated as
\begin{equation}
F_{q}^0(L)={\frac {1}{\beta^*(1-q)}}-{\frac {1}{\beta^*(1-q)}}C_q
\;,
\end{equation}
with
\[
C_q=\biggl\{
              \begin{array}{ll}
               ({\frac {L}{\delta_b (3-q)}}\sqrt{\frac {2 M\pi}{\beta^*(1-q)}}
{\frac {\Gamma({\frac {1}{1-q}})}
{\Gamma({\frac {1}{1-q}}+{\frac {1}{2}})}})^{1-q}, & \ \ \ \ \ q<1 \\
               ({\frac {L}{2 \delta_b}}\sqrt{\frac {2 M\pi}{\beta^*(q-1)}}
               {\frac {\Gamma({\frac {1}{q-1}}-{\frac {1}{2}})}
               {\Gamma({\frac {1}{q-1}})}})^{1-q}, & \ \ \ \ \ q>1
              \end{array}
\;.
\]

Since the above two systems are exactly solved, Ref.~\cite{5}
employed them to illustrate the variational method there. In the
present paper, we will use them to illustrate the VPA method. That
is, by regarding the Hamiltonian of the classical particle in the
box as $H_0(\mu)$ in eq.(10) and using the VPA scheme in last
section, we will calculate the GFE for the classical harmonic
oscillator up to the third order  and then make comparisons among
the results up to the various orders and the exact result for the
oscillator. Next, the present section will be divided into three
subsections. In subsection A, the truncated variational
perturbation expressions for the classical harmonic oscillator
will be given up to the third order . The values of $\mu$ in the
truncated expressions will be determined according to the PMS in
subsection B, and the approximated values of the GFE will be
calculated and compared with the exact result in subsection C.

\subsection{Truncated Expressions}
\label{tru}

Take the width $L$ of the box as the adjustable parameter $\mu$ in
our VPA scheme. Then, $H_I(\mu)$ in last section is now
\begin{eqnarray}
H_I(L)=\biggl\{
              \begin{array}{ll}
               {\frac {M\omega^2 x^2}{2}}, &  |x|<{\frac {L}{2}} \\
               {\frac {M\omega^2 x^2}{2}}-\infty, &  |x|\ge {\frac {L}{2}}
              \end{array}
              \;.
\end{eqnarray}
For the classical particle in the box, the speed $v_n=n
\delta_b/(M L)$ corresponds to the particle energy $E_n^{(0)}$. In
a state with the energy $E_n^{(0)}$, the particle moves in the box
back and forth at the constant speed $v_n$, and so the probability
of finding the particle in one of the directions near $x$
($|x|<L/2$) is $dx/(v_n T)$ with $T=2 L/v_n$ (except for $n=0$),
but zero out of the box. Thus, in the present case, the matrix
elements $H_{I,nm}$ appeared in Eqs.(20), (24) and (26) can be
straightforwardly calculated as
\begin{equation}
H_{I,nm}=\delta_{nm}2\int_{-L/2}^{L/2} {\frac {M\omega^2
x^2}{2}}{\frac {dx}{2 L}}=\delta_{nm} {\frac {M\omega^2L^2}{24}}
\;,
\end{equation}
where $\delta_{nm}=0$ for $m\not=n$ and $\delta_{nn}=1$.

Now we are at the position to calculate the truncated expressions
for $F_q$ of the classical harmonic oscillator up to the third
order  from Eqs.(13),(20),(24) and (26). The zeroth-order
expression is Eq.(31). At the first order, one can have, from
Eq.(20),
\begin{eqnarray}
F_q^{(1)}&=&\sum_n p_q^q(E_n^0)H_{I,\; nn}={\frac
{M\omega^2L^2}{24}}(Z_{q,\; 0})^{-q}\int_0^N[1-(1-q)\beta^* {\frac
{n^2\delta_b^2}{2 M L^2}}]^{q/(1-q)} dn
 \nonumber \\ &=&
 {\frac
{M\omega^2L^2}{24}}(Z_{q,\; 0})^{-q}\biggl \{
 \begin{array}{ll}
 {\frac {L}{2\delta_b}}\sqrt{{\frac {2 M}{(1-q)\beta^*}}}
                B({\frac {1}{2}},{\frac {1}{1-q}}) , & q<1
  \\ {\frac {L}{2\delta_b}}\sqrt{{\frac {2 M}{(q-1)\beta^*}}}
                B({\frac {1}{2}},{\frac {q+1}{2(q-1)}}) , & q>1
 \end{array}
 \nonumber \\ &=&
 {\frac {M \omega^2 L^2}{24}}{\frac {3-q}{2}} C_q \;.
\end{eqnarray}
The right hand side of last equation is identical to Eq.(12) in
Ref.~\cite{5}.

At the second order in $\epsilon$, Eq.(24) leads to
\begin{eqnarray}
F_q^{(2)} &=&\beta^* q Z_{q,\; 0}^{q-1}\biggl
[(<H_I>_q^0)^2-\sum_n
            (p_q(E_n^{(0)}))^{2q-1} (H_{I,\; nn})^2\biggr ]
  \nonumber \\ &=&
            \beta^* q Z_{q,\; 0}^{q-1}\biggl [(F_q^{(1)})^2-{\frac
{M^2\omega^4L^4}{(24)^2}}(Z_{q,\;
0})^{1-2q}\int_0^N[1-(1-q)\beta^* {\frac {n^2\delta_b^2}{2 M
L^2}}]^{(2q-1)/(1-q)} dn\biggr ]
  \nonumber \\ &=&
           \beta^* q Z_{q,\; 0}^{q-1}(F_q^{(1)})^2
         - \beta^* q Z_{q,\; 0}^{q-1}{\frac {M^2\omega^4 L^4}{(24)^2}}(Z_{q,\; 0})^{1-2q}\biggl \{
 \begin{array}{ll}
 {\frac {L}{2\delta_b}}\sqrt{{\frac {2 M}{(1-q)\beta^*}}}
                B({\frac {1}{2}},{\frac {q}{1-q}}) , & q<1
  \\ {\frac {L}{2\delta_b}}\sqrt{{\frac {2 M}{(q-1)\beta^*}}}
                B({\frac {1}{2}},{\frac {3q-1}{2(q-1)}}) , & q>1
 \end{array}
 \nonumber \\ &=&
             \beta^* {\frac {M^2\omega^4 L^4}{(24)^2}} {\frac {(q-3)(1-q)^2}{4}} C_q \;.
\end{eqnarray}

Finally, we consider the third order. Eq.(26), in the present
case, is reduced to
\begin{eqnarray}
F_q^{(3)} &=& {\beta^*}^2 q(q+1) Z_{q,\;
0}^{2q-2}(<H_I>_q^0)^3-3{\beta^*}^2q^2
           Z_{q,\; 0}^{2q-2}\sum_n (p_q(E_n^{(0)}))^{2q-1}(H_{I,\; nn})^2<H_I>_q^0
           \nonumber \\ &&
           +{\beta^*}^2 q(2q-1)Z_{q,\; 0}^{2q-2}\sum_n (p_q(E_n^{(0)}))^{3q-2}
           (H_{I,\; nn})^3
  \nonumber \\ &=&
          {\beta^*}^2 q(q+1) Z_{q,\; 0}^{2q-2}(F_q^{(1)})^3-3{\beta^*}^2q^2
           Z_{q,\; 0}^{-1}{\frac
            {M^2\omega^4L^4}{(24)^2}} F_q^{(1)} \int_0^N[1-(1-q)\beta^* {\frac
             {n^2\delta_b^2}{2 M L^2}}]^{{\frac {2q-1}{1-q}}} dn
  \nonumber  \\ &&
           +{\beta^*}^2 q(2q-1)Z_{q,\; 0}^{-q}{\frac
              {M^3\omega^6L^6}{(24)^3}}
            \int_0^N[1-(1-q)\beta^* {\frac
           {n^2\delta_b^2}{2 M L^2}}]^{{\frac {3q-2}{1-q}}} dn
 \nonumber \\ &=&
          {\beta^*}^2 q(q+1) Z_{q,\; 0}^{2q-2}(F_q^{(1)})^3-3{\beta^*}^2q^2
           Z_{q,\; 0}^{-1}{\frac
            {M^2\omega^4L^4}{(24)^2}} F_q^{(1)} \biggl \{
 \begin{array}{ll}
 {\frac {L}{2\delta_b}}\sqrt{{\frac {2 M}{(1-q)\beta^*}}}
                B({\frac {1}{2}},{\frac {q}{1-q}}) , & q<1
  \\ {\frac {L}{2\delta_b}}\sqrt{{\frac {2 M}{(q-1)\beta^*}}}
                B({\frac {1}{2}},{\frac {3q-1}{2(q-1)}}) , & q>1
 \end{array}
  \nonumber  \\ &&
           +{\beta^*}^2 q(2q-1)Z_{q,\; 0}^{-q}{\frac
              {M^3\omega^6L^6}{(24)^3}}\biggl \{
 \begin{array}{ll}
 {\frac {L}{2\delta_b}}\sqrt{{\frac {2 M}{(1-q)\beta^*}}}
                B({\frac {1}{2}},{\frac {2q-1}{1-q}}) , & q<1
  \\ {\frac {L}{2\delta_b}}\sqrt{{\frac {2 M}{(q-1)\beta^*}}}
                B({\frac {1}{2}},{\frac {5q-3}{2(q-1)}}) , & q>1
 \end{array}
 \nonumber \\ &=&
 {\beta^*}^2{\frac {M^3\omega^6L^6}{(24)^3}}{\frac
 {(1+q)(3-q)(q-1)^3}{8}}C_q \;.
\end{eqnarray}
Thus, collecting Eqs.(31),(34),(35) and (36), one can get
$F_q^{I}(L)$, $F_q^{II}(L)$ and $F_q^{III}(L)$, the truncated
expressions for $F_q$ of the classical oscillator at the first,
second and third orders. When $q<1$, the truncated expression for
$F_q$ at the third order  is
\begin{eqnarray}
F_{q,sup}^{III}(L)&=&{\frac {1}{\beta^*(1-q)}}+ \biggl({\frac
         {L}{\delta_b (3-q)}}\sqrt{\frac {2 M\pi}{\beta^*(1-q)}} {\frac
         {\Gamma({\frac {1}{1-q}})} {\Gamma({\frac {1}{1-q}}+{\frac
         {1}{2}})}}\biggr)^{1-q}\biggl[-{\frac {2}{\beta^*(1-q)(3-q)}}
  \nonumber \\ && \hspace*{1cm}
        +{\frac {M \omega^2 L^2}{24}}-\beta^* {\frac {M^2\omega^4 L^4}{(24)^2}}
        {\frac {(1-q)^2}{4}}+{\beta^*}^2 {\frac {M^3 \omega^6
        L^6}{(24)^4}}(q+1)(q-1)^3\biggr] \;,
\end{eqnarray}
and when $3>q>1$, it is
\begin{eqnarray}
F_{q,sub}^{III}(L)&=&{\frac {1}{\beta^*(1-q)}}+ \biggl(({\frac
          {L}{2 \delta_b}}\sqrt{\frac {2 M\pi}{\beta^*(q-1)}}
               {\frac {\Gamma({\frac {1}{q-1}}-{\frac {1}{2}})}
               {\Gamma({\frac {1}{q-1}})}})^{1-q}\biggr)^{1-q}\biggl[-{\frac {1}{\beta^*(1-q)}}
  \nonumber \\ &&\hspace*{2.5cm}
               +{\frac {M \omega^2 L^2}{24}}{\frac {(3-q)}{2}}-\beta^* {\frac {M^2\omega^4 L^4}{(24)^2}}
               {\frac {(3-q)(1-q)^2}{8}}
  \nonumber \\ &&\hspace*{2.5cm}
               +{\beta^*}^2 {\frac {M^3 \omega^6
                L^6}{(24)^4}}{\frac {(q+1)(q-1)^3(3-q)}{2}}\biggr] \;,
\end{eqnarray}
For the case of $q>3$, $F_{q,sub}^{III}(L)$ in Eq.(38) is
divergent. In Eqs.(37) and (38), all the terms in which the powers
in $\omega^2$ are lower than the second and the third powers give
the truncated expressions for $F_q$ at the first and second orders
, respectively. Next subsection, we will determine $L$ in them
according to the PMS to get approximations for $F_q$.

In the following subsections, when a numerical calculation is
performed, we will take $M=1, \omega=1, \delta_b=1/2$, and we will
consider only the case of $q<2$ which is the convergent range of
$F_q$ for the oscillator. ($\delta_b=1/2$ corresponds to the case
$h=1$ in Ref.~\cite{5}.)

\subsection{Determining $L$ }
\label{L}

$L$ must be positively real.

Up to the first order in $\epsilon$, for the cases of both
superextensiveness and subextensiveness,  Eq.(22) leads to the
equation
\begin{equation}
-{\frac {48}{\beta^*}} + M\omega^2 (-3 + q)^2 L^2=0 .
\end{equation}
To see whether $F_q^{I}(L)$ is most insensitive to $L$ near the
positive root of Eq.(39) or not, one can analyze varying property
of $F_q^{I}(L)$ as a function of $L$. For illustration, the
function $F_q^{I}(L)$ is depicted for $\beta^*=1$ and some values
of $q$ in Fig.1.
\begin{figure}[h]
\includegraphics{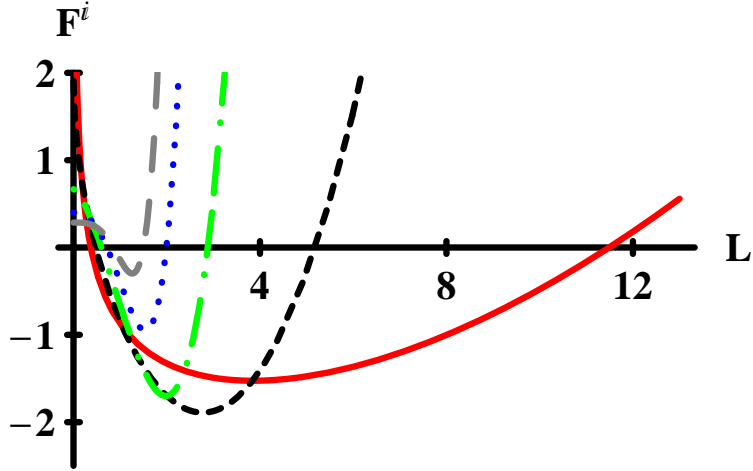}
\caption{\label{Fig.1} The dependence of $F_q^{I}(L)$ upon $L$ for
some values of $q$ and at $\beta^*=1$. The solid, dashed,
dot-dashed, dotted and long-dashed curves from right to left
correspond to $q=1.5, 0.5, -0.5, -1.5$ and $-2.5$, respectively.
Every curve in this figure has a minimum at a positive root for
$L$. }
\end{figure}
Fig.1 indicates that the positive root of Eq.(39) for $L$ makes
$F_q^{I}(L)$ reach a minimum and simultaneously be most
insensitive to $L$. So, we should choose this positive root as the
value of $L$
\begin{equation}
L^I={\frac {4}{3-q}}({\frac {3}{M\omega^2\beta^*}})^{1/2}  \;,
\end{equation}
which is identical to Eq.(13) in Ref.~\cite{5}. The minimum
$F_q^{I}(L^I)$ provides an upper limit for $F_q$. Thus, the VPA up
to the first order is really a variational method and it, in the
case of $q>0$, is consistent with the variational method in
Ref.~\cite{5} which is based on the generalized Bogoliubov
inequality, Eq.(9) in Ref.~\cite{5}.

Up to the second order , for the cases of both superextensiveness
and subextensiveness, Eq.(25) leads to the equation
\begin{equation}
4608 \beta^{*-2}-96L^2 M(q-3)^2 \beta^{*-1}\omega^2+L^4
M^2(q-1)^2(15-8q+q^2)\omega^4=0  \;.
\end{equation}
Last equation produces four roots for $L$. The roots are real only
when $-1.96239<q<1.62222$ and $3<q<5.34017$, and two of them are
positive as well as the other roots are negative. To choose an
appropriate root as $L$, $F_q^{II}(L)$, as a function of $L$, is
representatively illustrated for $\beta^*=1$ and some values of
$q$ in Fig.2.
\begin{figure}[h]
\includegraphics{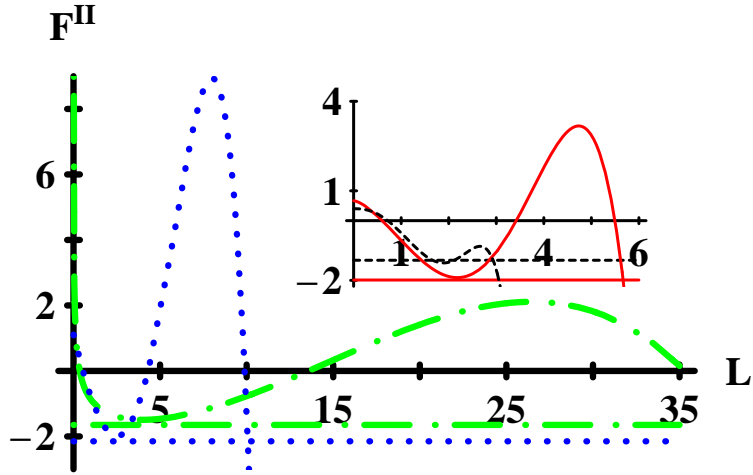}
\caption{\label{Fig.2} The dependence of $F_q^{II}(L)$ upon $L$
for some values of $q$ and at $\beta^*=1$. The dot-dashed and
dotted curves are for $q=1.25$ and $0.1$, respectively, and the
solid and dashed curves in the image are for $q=-0.5$ and $-1.5$,
respectively. Horizontal lines represent exact values of $F_q$,
and every pair of curve and horizontal line with an identical
line-type is for the same value of $q$. In the figure, every curve
has a minimum and a maximum. }
\end{figure}
In Fig.2, horizontal lines represent exact values of $F_q$, and
every pair of curve and horizontal line with an identical line
type is for the same value of $q$. From curves in Fig.2, one can
see that the smaller positive root makes $F_q^{II}(L)$ minimal
whereas the larger positive root makes $F_q^{II}(L)$ maximal.
Hence, $F_q^{II}(L)$ is more insensitive to $L$ near the smaller
positive root than near the other positive root, and so the
smaller root should be chosen as the value of $L$, $L^{II}$, for
$-1.96239<q<1.62222$.

For the other values of $q$, Eq.(25) cannot produce reasonable
values for $L$. In this case, the vanishing requirement for the
second derivative of $F_q^{II}(L)$ with respect to $L$ should be
considered and yields the following equation for $L$
\begin{eqnarray}
&&-14 L^4 M^2 q^4\omega^4 +
    L^4 M^2 q^5 \omega^4 +
    2 L^2 M q^2 \omega^2(384 \beta^{*-1} - 83 L^2 M \omega^2)
     \nonumber \\ && +
    12 L^2 M \omega^2 (144 \beta^{*-1} - 5 L^2 M \omega^2)
    +24 L^2 M q^3 \omega^2(-4 \beta^{*-1} + 3 L^2 M \omega^2)
    \nonumber \\ &&+
    q (4608 \beta^{*-2} - 2016 L^2 M \beta^{*-1} \omega^2 +
          167 L^4 M^2 \omega^4)=0 \;,
\end{eqnarray}
which is valid for the cases of both  superextensiveness and
subextensiveness. Eq.(42) produces four roots for $L$. There exist
real roots only when $-3.02227<q<3$ and $3.95406<q<5.601$, and for
$q<0$, two of them are positive as well as the other roots are
negative, and for $q>0$, only one positive root exists. To choose
an appropriate root as $L$ for the case of $-3.02227<q<-1.96239$,
$F_q^{II}(L)$, as a function of $L$, is typically illustrated for
$\beta^*=1$ and $q=-2$ in Fig.3. For completeness, we also depict
$F_q^{II}(L)$ for $q=1.8$ (which is involved in the range of
$1.62222<q<2$) in Fig.3 as an image.
\begin{figure}[h]
\includegraphics{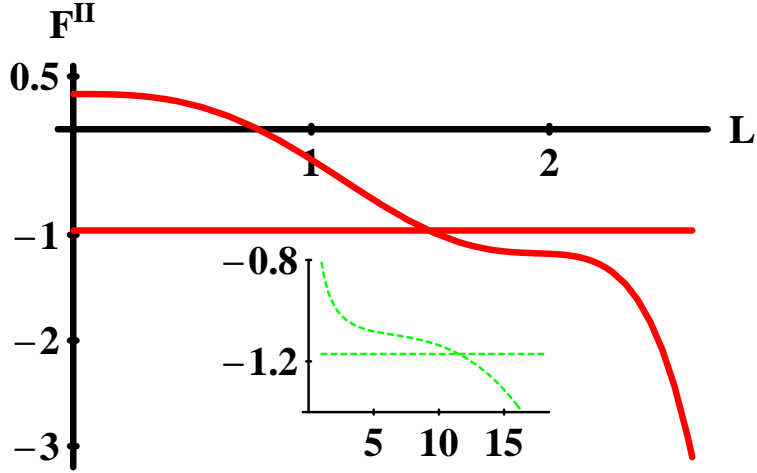}
\caption{\label{Fig.3} Similar to Fig.2, and the solid curve is
for $q=-2$ and the dashed curves in the image is for $q=1.8$. The
solid curve has two knees and the dashed line in the image has
only one knee.}
\end{figure}
In this figure, the solid curve for $q=-2$ has two knees which
correspond to the two positive roots of Eq.(42) for $L$: $1.13365$
and $1.94787$. Obviously, $F_q^{II}(L)$ is more insensitive to $L$
near the larger positive root than near the other positive root.
The same situation occurs for other negative values of $q$
($-3.02227<q<-1.96239$), and so the larger positive root of
Eq.(42) should be chosen as the value of $L$, $L^{II}$, for
$-3.02227<q<-1.96239$. Furthermore, as is illustrated in the image
of Fig.3, for the case of $2>q>1.62222$, $F_q^{II}(L)$ is most
insensitive to $L$ near the positive root of Eq.(42), and so the
positive root of Eq.(42) should be chosen as the value of $L$,
$L^{II}$, for $1.62222<q<2$. By the way, for the case of
$q<-3.02227$, there exists not any appropriate root of Eq.(41) or
Eq.(42) for $L$ and so the VPA method cannot provide an
approximation for $F_q$  up to the second order in this case.

Finally, we determine $L$ so as to approximate $F_q$ up to the
third order. For the cases of both  superextensiveness and
subextensiveness, Eq.(27) leads to the equation
\begin{eqnarray}
&&-663552\beta^{*-3} +
    13824L^2 M (-3 + q)^2\beta^{*-2}\omega^2 -
    144L^4 M^2(-1 + q)^2(15 - 8q +
          q^2)\beta^{*-1} \omega^4
    \nonumber \\ && \hspace*{2cm} +
    L^6 M^3(-1 + q)^3(21 + 11 q - 9 q^2 +
          q^3)\omega^6=0
\end{eqnarray}
Last equation produces six roots for $L$. Among the six roots,
there exist three positive roots for $-3<q<-1$, two positive roots
for $-1<q<1$ and only one positive root for $q<-3$, $1<q<3$ and
$q>7$. In Fig.4, $F_q^{III}(L)$, as a function of $L$, is
typically depicted for some values of $q$ and $\beta^*$.
\begin{figure}[h]
\includegraphics{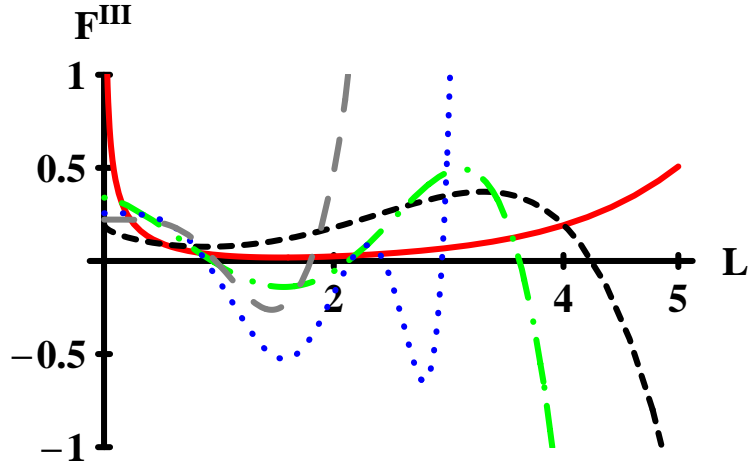}
\caption{\label{Fig.4} The dependence of $F_q^{III}(L)$ upon $L$
for some values of $q$ and $\beta^*$. The solid, short-dashed,
dot-dashed, dotted and long-dashed curves correspond to
$\{q,\beta^*\}=\{1.5,10\},\{0.5,10\},\{-0.5,1.9608\},\{-2.9,1\}$
and $\{-3.5,1\}$, respectively. The solid and the long-dashed
curves have a minimum, the short-dashed and the dot-dashed curves
have a minimum and a maximum, and the dotted curve has two minima
and one maximum. }
\end{figure}
In Fig.4, the dotted curve is for $\{q,\beta^*\}=\{-2.9,1\}$, and
has two minima and a maximum, which corresponds to the case of
having three positive roots. In this case, it is evident, from
Fig.4, that $F_q^{III}(L)$ is most insensitive to $L$ near the
smallest positive root of Eq.(43) (the curvature of the curve at
the smallest positive root is smaller than that at the largest
positive root), and so the smallest positive root should be chosen
as the value of $L$, $L^{III}$, for $-3<q<-1$. The short-dashed
and the dot-dashed curves are for $\{q,\beta^*\}=\{0.5,10\}$ and
$\{-0.5,1.9608\}$ and have a minimum at the smaller positive root
and a maximum at the other positive root. So, for the case of
having two positive roots, the smaller positive root is
appropriate for $L$ when $-1<q<1$. As for the cases of $q<-3$ and
$1<q<2$, the solid and the long-dashed curves indicate that
$F_q^{III}(L)$ reaches an extremum, the minimum at the positive
root, and so it can be taken as the value of $L$, $L^{III}$.

According to the PMS, we have determined $L$ for all cases which
we are interested in. In the above, analytical and numerical
discussions are done with Mathematica, and we do not list all
expressions of roots of Eqs.(41), (42) and (43) because they are
too lengthy. Note that although $L$ is determined as
$L^{I}$,$L^{II}$ and $L^{III}$ for approximating $F_q$ up to the
first, second and third orders, respectively, it is not meant that
$L$ is being expanded as a series.

\subsection{Generalized Free energy and Comparisons}
\label{com}

From analysis and results in last subsection, employing
Mathematica package, one can get, for various ranges of $q$, the
expressions of $F_q^{I}(L=L^I)$, $F_q^{II}(L=L^{II})$ and
$F_q^{III}(L=L^{III})$, the approximations of the GFE $F_q$ up to
the first, second and third orders. Regarding them as functions of
$t$ ($t=1/\beta^*$) and the non-extensiveness index $q$,
respectively, we can numerically calculate and compare them with
the exact $F_q$, Eq.(29).

The exact free energy, $F_q$ in Eq.(29), reaches a maximum at
$t=\delta_o (2-q)^{q/(q-1)}$ for any given value of $q<2$. As a
function of $q$, $F_q$ in Eq.(29) has a maximum for any $t<0.17$
or so, and, otherwise, reaches first a maximum and then a minimum
when $q$ increases up to $q=2$.

In Ref.~\cite{5}, the approximate GFE from variational method
there was considered for $0<q<2$, and when $t$ increases or when
$q$ decreases, the discrepancies between the approximate values
and the exact values become more and more evident. Here, we
consider the VPA GFE up to the first order, $F_q^{I}(L=L^I)$, for
all values of $q<2$, which with $0<q<2$ is identical to that in
Ref.~\cite{5}. Basically, $F_q^{I}(L=L^I)$, as a function of $t$
or $q$, mimics the feature of the exact free energy, and for a
given value of $t$, when $q$ approaches $2$ or is sufficiently
negative, the error $\Delta F\equiv F_q^{I}(L=L^I)-F_q$ is very
small whereas $\Delta F$ is not small for intermediate values of
$q$. Interestingly, the dependent feature of $\Delta F$ upon $q$
is similar to that of curvature of $F_q^{I}(L)$ at $L=L^I$,
$F''\equiv {\frac {d^2 F_q^{I}(L)}{d L^2}}|_{L=L^I}$, upon $q$
(For subextensiveness case, when $t$ is small, there exists not
such a similarity.). For an illustration of this similarity, in
Fig.5, we depict the dependence of $\Delta F$ (dashed curve) and
$F''$ (solid curve) upon $q$ at $t=3$.
\begin{figure}[h]
\includegraphics{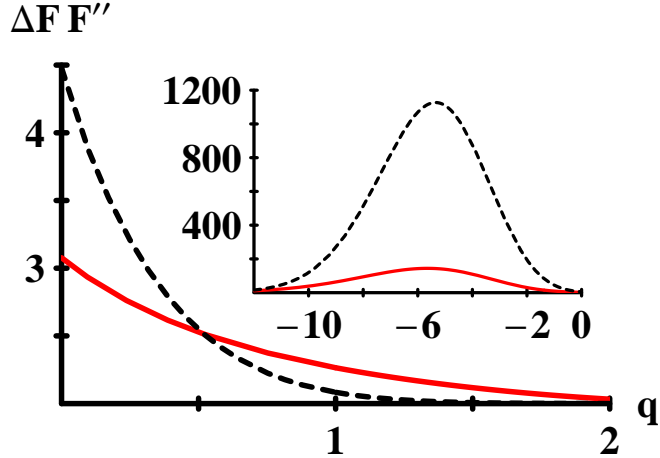}
\caption{\label{Fig.5} The dependence of $\Delta F$ (dashed curve)
and $F''$ (solid curve) upon $q$ at $t=3$. }
\end{figure}
In Fig.5, the image is drawn for the range of $-12<q<0$, and
evidently indicates that while the solid curve reaches the
maximum, the dashed curve also gets to its maximum. This
similarity can be understood from the PMS. Smaller the curvature
at $L=L^I$ is, more slowly $F_q^{I}(L)$ varies near $L=L^I$, and
so, closer $F_q^{I}(L=L^I)$ approaches $F_q$ according to the
spirit of the PMS, yielding the similarity.

The approximation of $F_q$ up to the second order,
$F_q^{II}(L=L^{II})$, improves the approximation of $F_q$ up to
the first order, $F_q^{I}(L=L^I)$, the variational result for the
range of $-3.02<q<2$ where $F_q^{II}(L=L^{II})$ makes senses, and
the approximation of $F_q$ up to the third order,
$F_q^{III}(L=L^{III})$, generally further improves the variational
result. For a comparison and an illustration, the exact and
various approximate results are shown at $t=1$ in Fig.6.
\begin{figure}[h]
\includegraphics{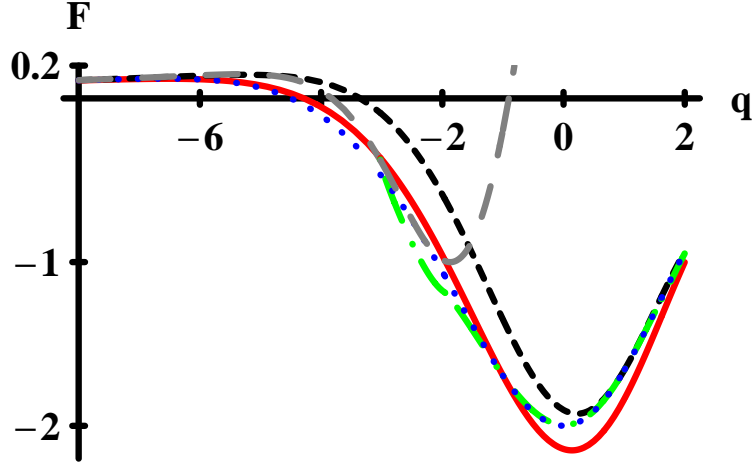}
\caption{\label{Fig.6} The dependence of $F_q$ and its
approximations up to various orders upon $q$ at $t=1$. The solid,
short-dashed, dot-dashed and dotted curves are the exact $F_q$ and
its approximations $F_q^{I}(L=L^I)$, $F_q^{II}(L=L^{II})$ and
$F_q^{III}(L=L^{III})$, respectively. The dot-dashed curve
interrupts at $q=-3.02$ or so, and almost coincides with the
dotted curve when $q>-1$ or so. The long-dashed curve is
$F_q^{III}(L=L^{I})$, and goes up quickly when $q>-2$ or so. }
\end{figure}
In Fig.6, the solid, short-dashed, dot-dashed and dotted curves
are $F_q$, $F_q^{I}(L=L^I)$, $F_q^{II}(L=L^{II})$ and
$F_q^{III}(L=L^{III})$, respectively (note that the dot-dashed
curve, $F_q^{II}(L=L^{II})$, interrupts at $q=-3.02$ or so, and
almost coincides with the dotted curve, $F_q^{III}(L=L^{III})$,
when $q>-1$ or so). Additionally, the long-dashed curve in Fig.6
is $F_q^{III}(L=L^{I})$, and obviously it is a bad approximation
for $F_q$ when $q>-2$ or so, albeit $F_q^{II}(L=L^{I})$ can
produce as a good approximation for $F_q$ as $F_q^{II}(L=L^{II})$
does. Fig.6 indicates that the improvement of $F_q^{II}(L=L^{II})$
to the variational result is substantial when $-3.02<q<0$.
Furthermore, the approximation of $F_q$ up to the third order
mimics the exact $F_q$ better than $F_q^{II}(L=L^{II})$, and
substantially improves the variational result for the range of
$-6<q<-3.02$ where $F_q^{II}(L=L^{II})$ is invalid. For the cases
of both $q<-6$ and $q>0$, $F_q^{II}(L=L^{II})$ and/or
$F_q^{III}(L=L^{III})$ only slightly improve the variational
result.(By the way, because $F_q$ and its various approximations
here vary slowly with $q$ near their maxima, the humps of those
curves in Fig.6 are not evident.)

For further and more clear illustration and comparison, we also
consider the dependence of the various approximations here upon
$t$. In Figs.7,8,9 and 10, $F_q$, $F_q^{I}(L=L^I)$,
$F_q^{II}(L=L^{II})$ and $F_q^{III}(L=L^{III})$ are depicted for
some typical values of $q$ as the solid, short-dashed, dot-dashed
and dotted curves, respectively, and, for the sake of clearness,
we redraw them for the range of $0<t<2$ in the image.
\begin{figure}[h]
\includegraphics{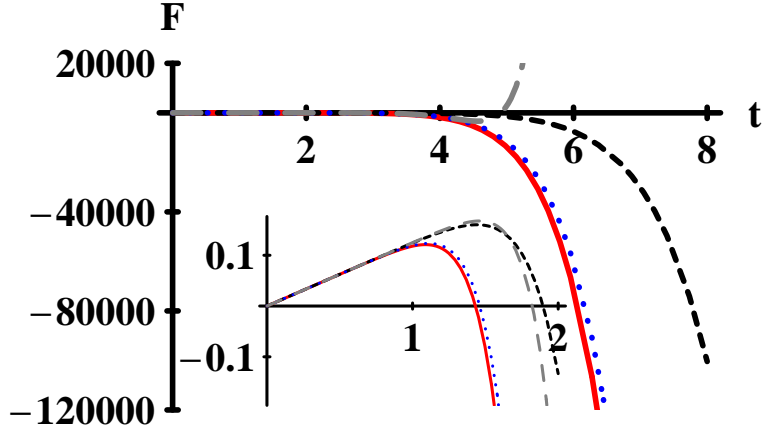}
\caption{\label{Fig.7} The dependence of $F_q$ and its
approximations up to various orders upon $t$ at $q=-7$. The solid,
short-dashed and dotted curves are the exact $F_q$ and its
approximations $F_q^{I}(L=L^I)$ and $F_q^{III}(L=L^{III})$,
respectively. The long-dashed curve is $F_q^{III}(L=L^{I})$, going
up when $t>5$ or so. They are redrawn for the range of $0<t<2$ in
the image.}
\end{figure}
\begin{figure}[h]
\includegraphics{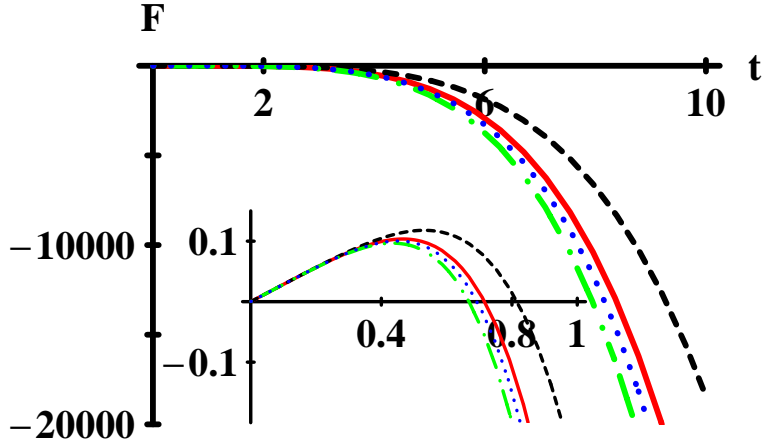}
\caption{\label{Fig.8} Similar to Fig.7, but $q=-2.5$, and the
dot-dashed is $F_q^{II}(L=L^{II})$. }
\end{figure}
\begin{figure}[h]
\includegraphics{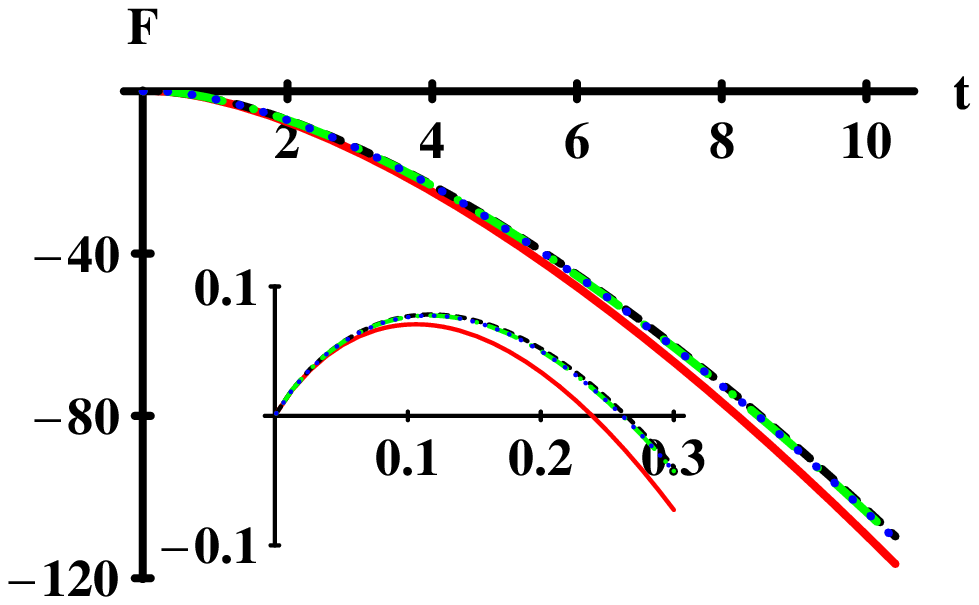}
\caption{\label{Fig.9}  Similar to Fig.7, but $q=0.5$ }
\end{figure}
\begin{figure}[h]
\includegraphics{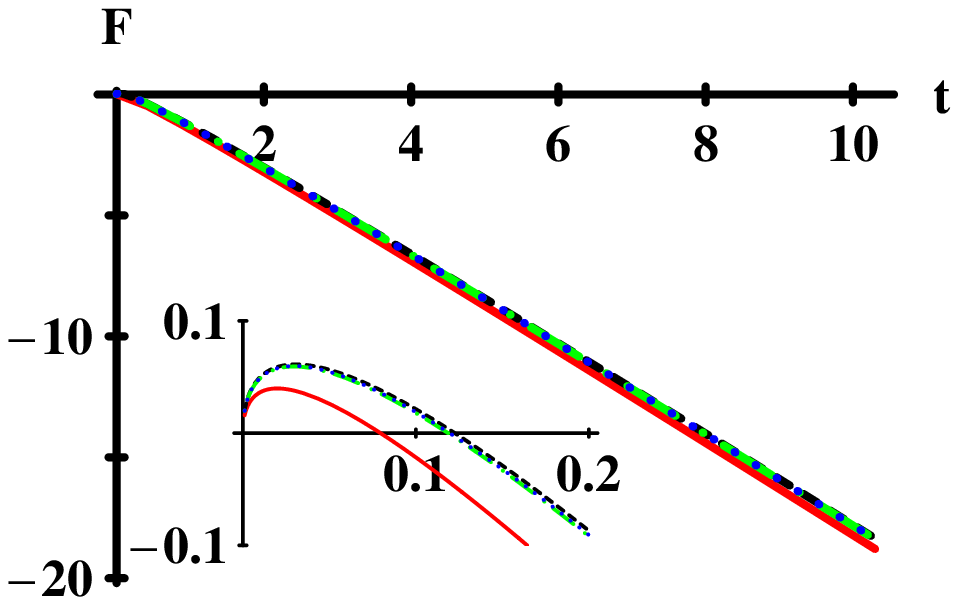}
\caption{\label{Fig.10}  Similar to Fig.7, but $q=1.5$}
\end{figure}

Fig.7 is drawn for $q=-7$, and so there is not a dot-dashed curve
owing to the invalidness of $F_q^{II}(L=L^{II})$ for this case. In
this figure, the long-dashed curve is $F_q^{III}(L=L^{I})$ which
goes up when $t>5$ or so, and is not given in Fig.8, 9 and 10
because it behaves too badly. Fig.7 indicates that although
$F_q^{II}(L=L^{II})$ does not exist, $F_q^{III}(L=L^{III})$
substantially improves the variational result and provide a better
approximation for $F_q$. Fig.8 is depicted for $q=-2.5$. In this
figure, although the dot-dashed and dotted curves are lower than
the solid curve, they indicate that in this case, the
approximations up to the first, second and third orders are
distinct from each other, and evidently the second- and
third-order approximations more closely approach the exact value
than the variational result. Fig.9 and 10 are drawn for $q=0.5$
and $1.5$, respectively, which was considered in Fig.1 in
Ref.~\cite{5}. For these cases, the short-dashed, dot-dashed and
dotted curves almost coincide and suggests that the approximations
up to the second and third orders provide only a very small
corrections to the variational results.

\section{Conclusion}
\label{5}

By considering the GFE for a system, this paper proposed a VPA
scheme for the generalized statistical mechanics based on the
Tsallis entropy. For approximating the GFE, we derived the
truncated expressions for $F_q$ up to the third order in the
variational perturbation expansion, and the classical harmonic
oscillator was considered in detail for an illustration. The model
investigation, albeit being a little academic \cite{12},
illustrates that the approximation up to the first order amounts
to a variational method and covers the variational method in
Ref.~\cite{5}, and the approximations up to the second and third
orders improve the variational result and tend to approach the
exact result.

Frankly, the variational perturbation expansion technique is
formally similar to the perturbation expansion in Ref.~\cite{5},
the work in the present paper is to introduce the variational
perturbation idea into the perturbation expansion and use the PMS
for determining the auxiliary parameter in the VPA scheme. It is
these revisions that make the variational perturbational
approximation method be non-perturbational, take the variational
result as the first-order approximation and systematically improve
the variational result. We believe that the investigation in the
present paper is useful, at least, for calculating the generalized
thermodynamical functions based on the Tsallis statistics.
Finally, we intend to point out that since path-integral formalism
has been developed \cite{1}, it is worth while developing VPA
method within the formalism of the generalized thermodynamics.

\end{document}